\documentclass[twocolumn,showpacs,preprintnumbers,superscriptaddress,prd]{revtex4}
\usepackage{amsmath,amssymb}
\usepackage{multirow}
\usepackage{graphicx}

\newcommand{\muq}{\mu_q}
\newcommand{\A}{\boldsymbol{A}}
\newcommand{\q}{\boldsymbol{q}}
\newcommand{\x}{\boldsymbol{x}}
\newcommand{\ms}{M_s}
\newcommand{\dmu}{\delta\mu}

\begin{document}

\preprint{RBRC-595}

\title{Characterizing the Larkin-Ovchinnikov-Fulde-Ferrel phase
induced by the chromomagnetic instability}
\author{Kenji Fukushima}
\affiliation{RIKEN BNL Research Center, Brookhaven National
 Laboratory, Upton, New York 11973, USA}

\begin{abstract}
  We discuss possible destinations from the chromomagnetic instability
  in color superconductors with Fermi surface mismatch $\delta\mu$.
  In the two-flavor superconducting (2SC) phase we calculate the
  effective potential for color vector potentials $\A_\alpha$ which
  are interpreted as the net momenta $\q$ of pairing in the
  Larkin-Ovchinnikov-Fulde-Ferrel (LOFF) phase.  When
  $1/\sqrt{2}<\delta\mu/\Delta<1$ where $\Delta$ is the gap energy,
  the effective potential suggests that the instability leads to a
  LOFF-like state which is characterized by color-rotated phase
  oscillations with small $\q$.  In the vicinity of
  $\delta\mu/\Delta=1/\sqrt{2}$ the magnitude of $\q$ continuously
  increases from zero as the effective potential has negative
  larger curvature at vanishing $\A_\alpha$ that is the Meissner mass
  squared.  In the gapless 2SC (g2SC) phase, in contrast, the
  effective potential has a minimum at
  $g\A_\alpha\sim\delta\mu\sim\Delta$ even when the negative Meissner
  mass squared is infinitesimally small.  Our results imply that the
  chromomagnetic instability found in the gapless phase drives the
  system toward the LOFF state with $\q\sim\delta\mu$.
\end{abstract}
\maketitle


     Quark matter has a rich phase structure in the high baryon or
quark density region.  In a decade we have witnessed tremendous
developments in theory, particularly in superconductivity of quark
matter~\cite{Rajagopal:2000wf}.  Color superconductivity is inevitable
from the Cooper instability in cold and dense quark matter.  In the
asymptotic density where the perturbative technique is applicable, the
color-flavor locked (CFL) phase~\cite{Alford:1998mk} where all quarks
are gapped is concluded from the first-principle calculations of
Quantum Chromodynamics (QCD).

     The lower density region we explore, the more complicated phase
possibilities we have to encounter.  The main reason why the situation
is perplexing at intermediate density is that a ``stress'' between
quarks which would form a Cooper pair is substantial when the quark
chemical potential, $\muq$, is comparable to the strange quark mass,
$\ms$.  Such an energy cost by the stress, or the Fermi energy
mismatch $\dmu$, is necessary to bind two quarks into a pair with zero
net momentum, $\q=0$.  The stress can be reduced by making a pair
between quarks sitting on different Fermi surfaces, which results in
$\q\neq0$.  If the energy gain by easing the stress is greater than
the kinetic energy loss coming from nonzero net momentum, the color
superconducting phase with $\q\neq0$ would be realized.  Since such a
state breaks rotational symmetry, this \textit{crystalline color
superconducting phase}~\cite{Alford:2000ze}, that is, a QCD analogue
of the Larkin-Ovchinnikov-Fulde-Ferrel (LOFF) phase~\cite{LOFF}, takes
a crystal structure~\cite{Bowers:2002xr}.

     There is another different possibility to consider while keeping
$\q=0$;  once $\dmu$ exceeds the gap energy, $\Delta$, the Cooper pair
tends to decay into two quarks. In other words the corresponding
quarks have the energy dispersion relation which is gapless.  Such a
phase is called the \textit{gapless superconducting
phase}~\cite{Gubankova:2003uj,Shovkovy:2003uu,Alford:2003fq}, that is,
a QCD analogue of the Sarma phase~\cite{Sarma}.  It would need a
careful comparison of energies to see which is favored in
reality~\cite{Casalbuoni:2005zp,Mannarelli:2006fy}.  Interestingly
enough, recently, these two different candidates, the crystalline and
gapless superconducting phases, have turned out to be closely related
through instability.

\begin{figure}
\includegraphics[width=6cm]{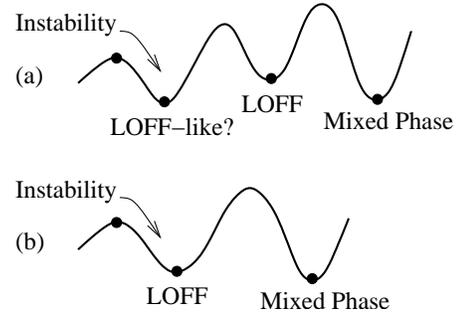}
\caption{Schematic energy landscape:  (a) The state falls from the
unstable gapless phase down toward a LOFF-like phase which is
separated from the LOFF phase (if it exists).  There are alternatives
such as the LOFF phase, mixed phase, and so on, which should be
energetically compared to the LOFF-like phase.  (b) The instability
directly leads to the LOFF phase.}
\label{fig:landscape}
\end{figure}

     The gapless superconducting phase is known to be unstable in fact
and has to give way to some other stable states.  In QCD the negative
(color) Meissner mass squared exhibits what is commonly referred to as
the chromomagnetic
instability~\cite{Huang:2004bg,Casalbuoni:2004tb,Alford:2005qw,Fukushima:2005cm}.
This is the central issue we address.  The chromomagnetic instability
is to be interpreted as instability toward the single plane-wave LOFF
state~\cite{Giannakis:2004pf}, as we will closely discuss later.  Also
it has been revealed that the two-flavor and three-flavor LOFF phases
are chromomagnetic stable~\cite{Giannakis:2005vw,Ciminale:2006sm} (see
also Ref.~\cite{Gorbar:2005tx}).  However, it does not necessarily
mean that the instability problem has already been resolved.   The
question we raise is as follows;  can we simply identify the
instability-induced state with the stable LOFF phase?  It is certain
that the instability tends to favor a LOFF-like state, but such a
LOFF-like state might exist separated from the LOFF phase (if it
exists) as sketched in Fig.~\ref{fig:landscape}(a), and then the
proposed stable LOFF phase is not the \textit{destination} from the
instability but one \textit{alternative} free from the instability
like a mixed phase.  Of course, it might be possible that the
instability is directly connected to the LOFF phase as sketched in
Fig.~\ref{fig:landscape}(b).

     To address this question we have to define the qualitative
difference between the LOFF and LOFF-like phases.  We shall
distinguish them by their characteristic wave numbers.  That is, if
the net momentum $\q$ is given as of order $\delta\mu$, then we regard
the system as going to the conventional LOFF state that has
$|\q|\simeq1.2\dmu$.  If $\q$ is small enough to be well separated
from $\q\sim\delta\mu$ inherent to the LOFF phase, we consider that
the system is then in the LOFF-like state.  For the purpose of
clarifying which situation of Figs.~\ref{fig:landscape}(a) and (b) is
more relevant, we will calculate the free energy as a function of
$\q$, or the color vector potential $\A_\alpha$ ($\alpha$ being the
adjoint color index) which is related to $\q$ through the covariant
derivative.

     As a preparation for our discussions we shall briefly look over
the color superconducting phases of our interest, the Meissner mass in
respective phases, and associated chromomagnetic instability.  The
predominant pairing is
\begin{equation}
 \Delta_\eta \sim \epsilon_{\eta ab}\epsilon_{\eta ij}
  \langle\bar{\psi}_{ai}\gamma_5\psi^C_{bj}\rangle
\label{eq:diquark}
\end{equation}
with $\psi^C=C\bar{\psi}^T$ and $a$, $b$ and $i$, $j$ being the color
and flavor indices, respectively.  Under this color-flavor locked
ansatz, $\Delta_1\neq0$, $\Delta_2\neq0$, and $\Delta_3\neq0$ defines
the CFL phase, while the two-flavor superconducting (2SC) phase has
$\Delta_3\neq0$ and $\Delta_1=\Delta_2=0$, that means only $ru$-$gd$
and $rd$-$gu$ quarks make a pair.  The gapless 2SC and CFL phases
(abbreviated as the g2SC and gCFL phases) occur when
$\dmu\approx\mu_e/2>\dmu_c^{\text{g}}=\Delta_3$ and
$\dmu\approx\ms^2/2\muq>\dmu_c^{\text{g}}=\Delta_1$, respectively,
where $\mu_e$ is the electron chemical potential.  In the single
plane-wave LOFF ansatz the gap parameters are augmented as
\begin{equation}
 \Delta_\eta \;\xrightarrow{\text{LOFF}}\;
  \exp[-2i\q_\eta\cdot\x]\,\Delta_\eta.
\label{eq:LOFF}
\end{equation}

     The Meissner mass is the screening mass for transverse gauge
fields.  The individual mass is a quantity dependent on the gauge
choice;  we  can arbitrarily shuffle eight gluon fields
$\A_1,\dots,\A_8$ by a gauge rotation.  It should be noted, however,
that the choice of the diquark condensate (\ref{eq:diquark}) specifies
a gauge direction and then the Meissner mass is uniquely determined.
The finite Meissner mass arises associated with spontaneous symmetry
breaking and the Higgs-Anderson mechanism in
superconductors~\cite{Rischke:2000qz}.

     In the 2SC phase $\A_1$, $\A_2$, and $\A_3$ remain massless and
the rest $\A_4,\dots,\A_8$ earn a finite Meissner mass.  The
electromagnetic field $\A_\gamma$ has mixing with $\A_8$ which ends up
with two eigen-fields $\A_{\tilde{8}}$ and $\A_{\tilde{\gamma}}$.  The
system has symmetry among $\A_4,\dots,\A_7$ and thus the resulting
Meissner mass is common for them, while $\A_{\tilde{8}}$ has a
different Meissner mass. Because the modified (i.e.\ color-mixed)
electromagnetic U(1) symmetry is unbroken, $\A_{\tilde{\gamma}}$ stays
massless.  It is known in the CFL phase at $\ms=0$, on the other hand,
that all eight gluons $\A_1,\dots,\A_8$ have a common and nonvanishing
Meissner mass.  In the presence of mixing with $\A_\gamma$ the massive
eigen-field $\A_{\tilde{8}}$ pulls away from the others and
$\A_{\tilde{\gamma}}$ is massless.

     In the two-flavor case at finite $\dmu$ away from the ideal 2SC
phase, there are still only two independent Meissner
masses~\cite{Huang:2004bg}; one is for $\A_4,\dots,\A_7$ and the other
is for $\A_{\tilde{8}}$.  The Meissner mass squared for
$\A_4,\dots,\A_7$ becomes negative (i.e.\ the Meissner mass is
imaginary) for $\dmu>\dmu_c^{\text{4-7}}=\dmu_c^{\text{g}}/\sqrt{2}$,
that means an instability occurs not only in the g2SC phase
($\dmu>\dmu_c^{\text{g}}$) but in the 2SC phase
($\dmu_c^{\text{4-7}}<\dmu<\dmu_c^{\text{g}}$) also.  The Meissner
mass squared for $\A_{\tilde{8}}$ is negatively divergent at the
gapless onset, $\dmu=\dmu_c^{\text{g}}$, and remains on negative in
the entire g2SC side.

    The three-flavor case with finite $\ms$ has a more complicated
pattern and there are five independent Meissner
masses~\cite{Fukushima:2005cm}.  With nonzero $\ms$ one should take
account of mixing among $\A_3$, $\A_8$, and $\A_\gamma$ properly, from
which two massive eigen-fields, $\A_{\tilde{3}}$ and $\A_{\tilde{8}}$,
and one massless $\A_{\tilde{\gamma}}$ result.  The Meissner mass is
degenerated for $\A_1$ and $\A_2$ due to symmetry, and so is for
$\A_4$ and $\A_5$, and for $\A_6$ and $\A_7$.  No instability takes
place until the system reaches the gCFL phase.  At
$\dmu=\dmu_c^{\text{g}}$ negatively divergent Meissner masses squared
appear for $\A_1$-$\A_2$ and for $\A_{\tilde{8}}$.  As for
$\A_4$-$\A_5$ and $\A_6$-$\A_7$, when $\dmu$ gets larger than critical
values $\dmu_c^{\text{4-5}}$ and $\dmu_c^{\text{6-7}}$ respectively,
they eventually have negative Meissner masses squared, which is
presumably related to the instability in the two-flavor calculation.
We shall summarize the instability patterns in
Table~\ref{tab:instability}.

\begin{table}
\begin{tabular}{c@{\hspace{2mm}}c@{\hspace{3mm}}c}
 \hline
  & two-flavor & three-flavor \\
 \hline
  $\A_1$, $\A_2$ & massless & unstable $\dmu>\dmu_c^{\text{g}}$ \\
  $\A_{\tilde{3}}$ & massless & stable \\
  $\A_4$, $\A_5$ &
   \multirow{2}{*}{$\bigg\}\;$unstable $\dmu>\dmu_c^{\text{4-7}}$}
   & unstable $\dmu>\dmu_c^{\text{4-5}}$ \\
  $\A_6$, $\A_7$ & & unstable $\dmu>\dmu_c^{\text{6-7}}$ \\
  $\A_{\tilde{8}}$ & unstable $\dmu>\dmu_c^{\text{g}}$
   & unstable $\dmu>\dmu_c^{\text{g}}$ \\
  $\A_{\tilde{\gamma}}$ & massless & massless \\
 \hline\\
\end{tabular}
\caption{Chromomagnetic instability for each gluon in the two-flavor
  and three-flavor cases at zero temperature.  Here
  $\dmu_c^{\text{4-7}}=\dmu_c^{\text{g}}/\sqrt{2}$ and
  $\dmu_c^{\text{4-5}}\simeq\dmu_c^{\text{6-7}}$ has been numerically
  estimated as $\sim 2.3\,\dmu_c^{\text{g}}$ in
  Ref.~\cite{Fukushima:2005cm}.}
\label{tab:instability}
\end{table}


     Now let us consider what the negative Meissner mass squared
signifies.  It is a textbook knowledge that in the $\phi^4$-theory,
for the simplest example, a nonzero expectation value
$\langle\phi\rangle\neq0$ grows when the screening mass squared for
$\phi$ is negative.  In the language of the effective potential the
negative mass squared means that a state lies in a maximum of the
potential and a true ground state should exist somewhere down away
from $\langle\phi\rangle=0$.  Therefore it is quite natural to
anticipate that the chromomagnetic instability is cured by nonzero
color vector potentials $\langle\A_\alpha\rangle$.  Actually the
Meissner mass squared is the coefficient of the quadratic terms in the
kinetic energy expanded in $\A_\alpha$,
\begin{equation}
 m_{M\alpha\beta}^2 = \frac{1}{3}\frac{\partial^2\Omega_{\text{kin}}}
  {\partial A_\alpha^i \partial A_\beta^i}\biggr|_{A=0},
\end{equation}
where the kinetic energy term takes a form of
\begin{align}
 &\Omega_{\text{kin}}[\Delta,A] \notag\\
 &= \kappa^{\eta\sigma}_{\eta'\sigma'}\bigl[(\partial^i
  \delta_{\eta\eta'}\!+igA^{\ast i}_{\eta\eta'})\Delta^\ast_{\eta'}
  \bigr]
  \bigl[(\partial^i\delta_{\sigma\sigma'}\!-igA^i_{\sigma\sigma'})
  \Delta_{\sigma'}\bigr] \notag\\
 &\qquad + \text{higher-order terms in $A$}
\label{eq:kinetic}
\end{align}
due to symmetry.  In the following discussions we shall ignore
mixing with $\A_\gamma$ for simplicity.  Here we should remark that
the stiffness parameter $\kappa$ depends on the ``flavor'' indices
$\eta'$ and $\sigma'$ as well as the ``color'' indices $\eta$ and
$\sigma$.  This assignment is understood from that $\eta$ of
$\Delta_\eta$ contains the information on both color and flavor as
fixed in (\ref{eq:diquark}).  The covariant derivative acting on a
color-triplet rotates $\eta'$ into $\eta$, while the flavor is intact
as $\eta'$.

     In the two-flavor case only $\kappa^{\eta\sigma}_{33}$ is
relevant and we can forget about ``flavor'', as parametrized in
Refs.~\cite{Iida:2006,Fukushima:2005fh}.  Then the chromomagnetic
instability with two independent Meissner masses in this case can be
expressed by a combination of two parameters
$\kappa^{(1)}$ and $\kappa^{(2)}$ where
$\kappa^{\eta\sigma}_{33}=\kappa^{(1)}\delta_{\eta\sigma}+\kappa^{(2)}\Delta_\eta\Delta^\ast_\sigma$
which makes $\Omega_{\text{kin}}$ a color-singlet.  The single
plane-wave LOFF state characterized by (\ref{eq:LOFF}) with only
$\Delta_3$ nonvanishing is sensitive to
$\kappa^{33}_{33}=\kappa^{(1)}+\kappa^{(2)}|\Delta_3|^2$ and thus such
a gap parameter ansatz cannot separate two distinct instabilities for
$\A_8$ with $m_{M88}^2\propto\kappa^{33}_{33}$ and for
$\A_4,\dots,\A_7$ with $m_{M4\mbox{-}7}^2\propto\kappa^{(1)}$ if they
coexist.

     It is intriguing to look into  the three-flavor case next.  The
stiffness parameter can be decomposed as 
\begin{equation}
 \kappa^{\eta\sigma}_{\eta'\sigma'}=\kappa^\lambda_{\text{off}}
  |\epsilon^{\lambda\eta\eta'}|\delta_{\eta\sigma}
  \delta_{\eta'\sigma'}
 +\kappa^{\eta\sigma}_{\text{diag}}\delta_{\eta\eta'}
  \delta_{\sigma\sigma'}.
\end{equation}
This decomposition is justified by the color and flavor structure in
the quark one-loop calculations~\cite{Fukushima:2005cm}.  Then the
first term involving $\kappa^\lambda_{\text{off}}$ is relevant to the
Meissner mass squared for the color-off-diagonal gluons;
\begin{equation}
 \begin{split}
  m_{M1\mbox{-}2}^2 \propto \kappa^3_{\text{off}}
   (|\Delta_1|^2+|\Delta_2|^2),\\
  m_{M4\mbox{-}5}^2 \propto \kappa^2_{\text{off}}
   (|\Delta_3|^2+|\Delta_1|^2),\\
  m_{M6\mbox{-}7}^2 \propto \kappa^1_{\text{off}}
   (|\Delta_2|^2+|\Delta_3|^2).
 \end{split}
\end{equation}
The Meissner mass squared for the diagonal gluons, on the other hand,
comes from the second term involving
$\kappa^{\eta\sigma}_{\text{diag}}$, that is, $m_{M33}^2$,
$m_{M38}^2$, and $m_{M88}^2$ are written as a linear combination of
six components of the symmetric $3\times3$ matrix
$\kappa^{\eta\sigma}_{\text{diag}}$.  We note that, when the ansatz
(\ref{eq:LOFF}) is substituted into (\ref{eq:kinetic}), the
instability toward finite $\q_\eta$ does not reflect the information
of $\kappa^\lambda_{\text{off}}$, that is,
$\partial^2\Omega_{\text{kin}}/\partial q_\eta\partial q_\sigma\propto
\kappa^{\eta\sigma}_{\text{diag}}$.

     From (\ref{eq:kinetic}) we can immediately understand how the
chromomagnetic instability is generally transformed to the instability
toward the single plane-wave LOFF phase, which has been shown for
$\A_8$ in explicit calculations in two-flavor quark matter in
Ref.~\cite{Giannakis:2004pf}.  The point is that the gauge fields in
the quark sector appear only in the covariant derivative, so that they
can be absorbed as a phase factor of the gap parameters.

     Now we assume that we have an instability only for $\A_8$.  The
rotational symmetry is broken and we choose the $n$-direction in
three-dimensional spatial space in which $\A_8$ acquires an
expectation value.  Then the covariant derivative is equivalently
rewritten as
\begin{equation}
 \begin{split}
 &\bigl[\partial^i\delta_{\eta\eta'}-\delta^{in}igA^n_8
  (t^8)_{\eta\eta'}\bigr]\Delta_{\eta'} \\
 & =\exp[igt^8 \A_8\cdot\x]_{\eta\eta'}\, \partial^i \bigl\{
  \exp[-igt^8\A_8\cdot\x]_{\eta'\eta^{\prime\prime}}
  \Delta_{\eta^{\prime\prime}}\bigr\},
 \end{split}
\end{equation}
where $t_\alpha$'s are the color group generators in the fundamental
representation.  The color rotation results in
\begin{equation}
 \exp[-igt^8\A_8\cdot\x]\cdot \Delta = \left(
  \begin{array}{c}
   \exp[-\frac{ig}{2\sqrt{3}}\A_8\cdot\x]\; \Delta_1 \\
   \exp[-\frac{ig}{2\sqrt{3}}\A_8\cdot\x]\; \Delta_2 \\
   \exp[+\frac{ig}{\sqrt{3}}\A_8\cdot\x]\; \Delta_3
  \end{array} \right),
\label{eq:act_A8}
\end{equation}
which is nothing but the diquark condensate peculiar to the
three-flavor single plane-wave LOFF state.  [We assumed that
$\A_\alpha$ is a constant, but the generalization to inhomogeneous
$\A_\alpha(\x)$~\cite{Gorbar:2006up} is easy; the exponential part is
then the Wilson line.]  From the above rewriting, it is apparent that
the non-LOFF (ordinary) superconducting phase with a color vector
potential $\A_8$ is equivalent to the LOFF phase whose spatial
oscillation is characterized by $\A_8$ with no vector potential.  Of
course, this general argument works in the two-flavor case as well;
$\Delta_1=\Delta_2=0$ and a phase factor emerges for $\Delta_3$ alone,
so one could interpret such an overall phase as associated with the
baryon number~\cite{Huang:2005pv}, though such an interpretation has
only a limited meaning.

     One has to be careful when this argument is applied for the
off-diagonal gluons, $\A_1$, $\A_2$, $\A_4$, $\A_5$, $\A_6$, and
$\A_7$.  For instance, if the instability occurs in $\A_4$, then the
phase factor is no longer in the form of the single plane-wave.  In
the same way as in the previous case we have
\begin{equation}
 \begin{split}
 &\exp[-igt^4\A_4\cdot\x]\cdot \Delta \\
 & = \left(
  \begin{array}{c}
   -i\sin[\frac{g}{2}\A_4\cdot\x]\; \Delta_3
    +\cos[\frac{g}{2}\A_4\cdot\x]\; \Delta_1 \\
   \Delta_2 \\
   \cos[\frac{g}{2}\A_4\cdot\x]\; \Delta_3
    -i\sin[\frac{g}{2}\A_4\cdot\x]\; \Delta_1
 \end{array} \right).
 \end{split}
\label{eq:act_A4}
\end{equation}
This represents not a single but rather multiple plane-wave LOFF
state, or \textit{color-rotated} single plane-wave LOFF.  In the
two-flavor case we keep $\Delta_3$ alone and then, interestingly,
(\ref{eq:act_A4}) indicates that we definitely need to have not only
the third component $\cos[\frac{g}{2}\A_4\cdot\x]\,\Delta_3$ but also
the first component $-i\sin[\frac{g}{2}\A_4\cdot\x]\,\Delta_3$ which is
not considered at all in the conventional two-flavor treatment.  Our
analysis agrees with the conclusion of Ref.~\cite{Gorbar:2005tx} that
the single plane-wave LOFF state would still have instability for the
off-diagonal gluons.


     From the discussions so far we can establish the
\textit{qualitative} (apart from a color rotation) correspondence
between the vector potentials and the net momenta of the single
plane-wave ansatz as
\begin{equation}
 \frac{g\A_8}{2\sqrt{3}} \longleftrightarrow 2\q,\quad
 \frac{g\A_4}{2} \longleftrightarrow 2\q,
\end{equation}
Now we shall estimate the magnitude of characteristic $\q$ as a result
of the instability using the above relations.

\begin{figure}
\includegraphics[width=7cm]{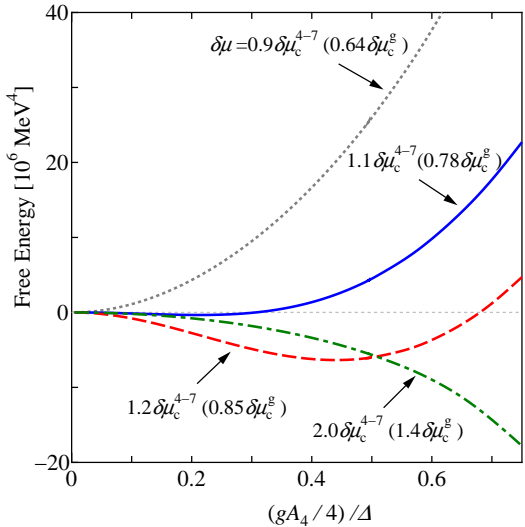}
\caption{Free energy difference from the energy without vector
potentials in two-flavor quark matter as a function of $\A_4$.  The
critical $\dmu$ for $\A_4$ is
$\dmu_c^{\text{4-7}}=\dmu_c^{\text{g}}/\sqrt{2}$ and the g2SC phase
occurs at $\dmu=\dmu_c^{\text{g}}$.}
\label{fig:a4} 
\end{figure}

\begin{figure}
\includegraphics[width=7cm]{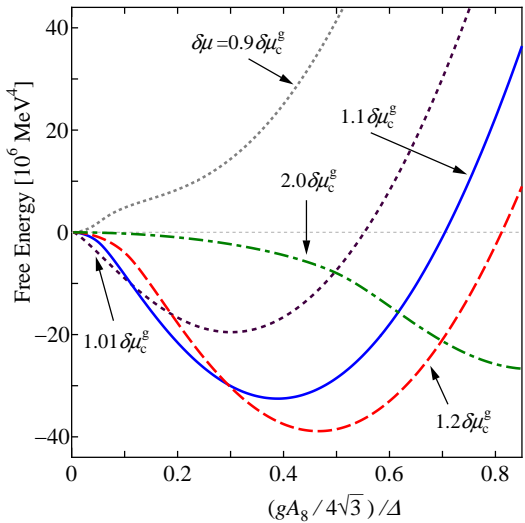}
\caption{Free energy difference from the energy without vector
potentials in two-flavor quark matter as a function of $\A_8$ in the
unit of $\Delta$.}
\label{fig:a8}
\end{figure}

     For that purpose we need to know the higher-order terms in
$\A_\alpha$ in the expansion (\ref{eq:kinetic}).  As we will explain
shortly, however, such an expansion is no longer valid in the gapless
phase.  Thus we must evaluate the $\A_\alpha$-dependent part of the
free energy without expansion.  For simplicity we will limit our
discussions only to the two-flavor calculations from now on.

     We write down the $48\times48$ (two-flavors, three-colors,
two-spins, particle-antiparticle, and two-Nambu-Gorkov-doublers)
quasi-quark propagator with either $\A_4$ (that we arbitrarily chose
among $\A_4,\dots\A_7$) or $\A_8$ and calculate the quasi-quark energy
$\epsilon(\boldsymbol{p})$ which depends on the momentum angle to the
vector potential, i.e., $\boldsymbol{p}\cdot\A_\alpha$.  The free
energy is available as integration of the sum over all 48
$|\epsilon(\boldsymbol{p})|$'s with respect to $\boldsymbol{p}$.  We
regulate the momentum integration by the ultraviolet cutoff
$\Lambda=1\,\text{GeV}$ and subtract the free energy at
$\Delta=\dmu=0$ to get rid of the cutoff artifact.  The gap parameter
is fixed at $\Delta=100\,\text{MeV}$.  It should be noted that the
analytical formulae utilized in the two-flavor LOFF
calculations~\cite{Giannakis:2005vw,Gorbar:2005tx} do not work for
$\A_4$ which is color off-diagonal.

     We present the numerical results in Figs.~\ref{fig:a4} and
\ref{fig:a8}.  The potential curvature at $\A_\alpha=0$ corresponds to
the Meissner screening mass squared.  In Fig.~\ref{fig:a4} the
Meissner mass is real finite for $\dmu<\dmu_c^{\text{4-7}}$, while the
origin $\A_4=0$ becomes unstable when $\dmu>\dmu_c^{\text{4-7}}$, as
is manifest from the results at $\dmu=0.9\dmu_c^{\text{4-7}}$ (dotted
curve) and $\dmu=1.1\dmu_c^{\text{4-7}}$ (solid curve).  This
instability occurs continuously and we can see from the
$\dmu=1.2\dmu_c^{\text{4-7}}$ results (dashed curve) that the expected
$\A_4$ grows as $\dmu$ approaches $\dmu_c^{\text{g}}$.  It is
known~\cite{Huang:2004bg} that the negative Meissner mass for $\A_4$
becomes small again when $\dmu$ is larger than $\dmu_c^{\text{g}}$.
Certainly our calculations for
$\dmu=2.0\dmu_c^{\text{4-7}}=1.4\dmu_c^{\text{g}}$ (dot-dashed curve)
result in smaller potential curvature and thus smaller Meissner mass
than those for $\dmu=1.2\dmu_c^{\text{4-7}}$.  Nevertheless, the
expected $\A_4$ is \textit{larger} and we find a potential minimum at
$g|\A_4|/4\simeq 1.39\Delta=0.98\dmu$.  In this way the results for
$\dmu>\dmu_c^{\text{g}}$ make a sharp contrast to the nature of the
instability for $\dmu\sim\dmu_c^{\text{4-7}}<\dmu_c^{\text{g}}$.  In
the gapless region where $\dmu>\dmu_c^{\text{g}}$, the expected
$g|\A_4|/4$ (and thus $\q$) is of order $\dmu$ \textit{however small
the potential curvature (Meissner mass squared) is}.

     The same observation is apparent also in Fig.~\ref{fig:a8}.  The
Meissner mass squared for $\A_8$ is negative divergent at
$\dmu=\dmu_c^{\text{g}}$, meaning that the potential has a cusp at
$\A_8=0$ then, which is confirmed in our results as seen at
$\dmu=1.01\dmu_c^{\text{g}}$ (short-dashed curve).  When $\dmu$ pulls
away from the onset value, the negative Meissner mass squared becomes
smaller, and at the same time, $\A_8$ acquires a larger expectation
value of order $\dmu$ again.

\begin{figure}
\includegraphics[width=7cm]{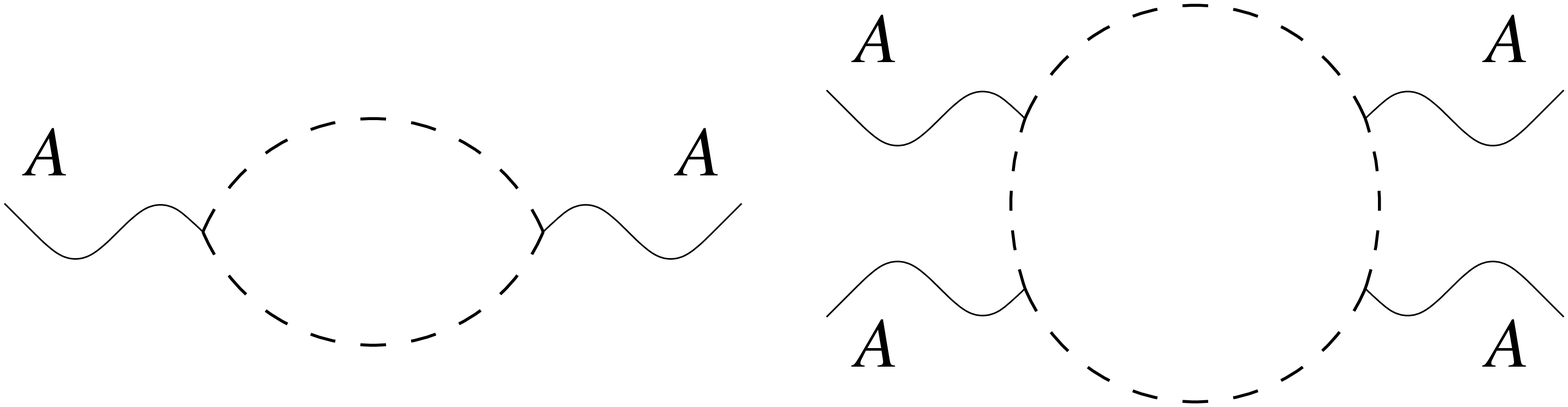}
\caption{Diagrammatic expansion of the free energy in terms of
$g\langle\A_\alpha\rangle/\epsilon(\boldsymbol{p})$ where the quark
energy $\epsilon(\boldsymbol{p})$ comes from the quark
propagator.}
\label{fig:expansion}
\end{figure}

     We would emphasize that these findings are unexpected results;
if the Ginzburg-Landau expansion of (\ref{eq:kinetic}) works with a
positive definite quartic term in $\A_\alpha$, an infinitesimal
negative $\kappa$ (potential curvature) simply leads to an
infinitesimal $\A_\alpha$. Therefore, our results imply that not only
the quadratic term but also quartic and even higher-order terms are
significantly affected by gapless quarks when
$\dmu>\dmu_c^{\text{g}}$.  The reason why the Ginzburg-Landau
expansion breaks down can be understood in a diagrammatic way.

     Figure~\ref{fig:expansion} shows an example of the diagrammatic
expansion of the free energy in terms of $\A_\alpha$.  The
dimensionless expansion parameter is obviously
$g\langle\A_\alpha\rangle/\epsilon(\boldsymbol{p})$ where
$\epsilon(\boldsymbol{p})$ is the quark energy stemming from the
propagator.  Therefore such an expansion is no longer legitimate once
gapless quarks whose $\epsilon(\boldsymbol{p})$ can become vanishingly
small enter the loop.  In other words, in the gapless phase, the
Meissner mass squared is far from informative on the true ground
state;  the smallness of the Meissner mass squared does not mean the
weakness of the instability.

     It should be noted that our potential analysis neither preserves
neutrality nor solves the equations of motion.  As long as $\A_\alpha$
is small, the coupling between $\A_\alpha$ and the other parameters
such as the chemical potentials and the gap parameters is small due to
approximate rotational symmetry.  We can thus expect that the free
energy we estimated would not be modified significantly by neutrality
and the condensation energy in the region where $\A_\alpha$ is small.
This is not the case, however, once $\A_\alpha\sim\dmu$ develops.
Hence, strictly speaking, we cannot say anything about the exact
location of the potential minimum, but we can at least insist that
there is no stable LOFF-like state with small $\q$ directly resulting
from the chromomagnetic instability in the gapless phase even when the
negative Meissner mass squared is tiny.

     Our results in the two-flavor case are not direct evidence but
suggestive;  we can anticipate that the situation in
Fig.~\ref{fig:landscape}(a) is realized for the off-diagonal gluons at
$\dmu_c^{\text{4-7}}<\dmu\ll\dmu_c^{\text{g}}$, while the situation in
Fig.~\ref{fig:landscape}(b) is likely to be the case for
$\dmu\gtrsim\dmu_c^{\text{g}}$.  In the three-flavor case we
conjecture that the instability picture is close to
Fig.~\ref{fig:landscape}(b) since the chromomagnetic instability then
occurs only in the gapless region of $\dmu$.  These are our central
conclusions derived from the numerical results.


     Finally let us comment on the possibility of coexistence of both
$\A_4$ and $\A_8$ in the two-flavor case.  We shall call such a state
the \textit{gluonic phase}~\cite{Gorbar:2005rx}.  One
should be careful about the terminology not to fall in a mere
interpretation;  we would use the nomenclature, the gluonic phase,
differently from the original usage in Ref.~\cite{Gorbar:2005rx}, but
to mean a state in which all $\A_\alpha$ in the covariant derivative
cannot be simultaneously removed by any gauge rotation of
$\Delta_\eta$.  Thus one can uniquely define the gluonic phase in a
way distinct from the LOFF and LOFF-like states.

     If $\A_4$ and $\A_8$ are not parallel, $A_4^1$ and $A_8^2$ for
instance, then we cannot find an appropriate gauge rotation
$\Delta\to V\Delta$ to eliminate them simultaneously.  Namely, the
gauge rotation matrix $V$ satisfying
$V^\dagger\partial^1 V=igA_4^1 t^4$ and
$V^\dagger\partial^2 V=igA_8^2 t^8$ does not exist if $V$ is assumed
not to have any singularity.

     The gluonic phase has one more significant difference from the
LOFF state besides the covariant derivative;  it has nonvanishing
chromomagnetic field.  For our example $A_4^1$ and $A_8^2$ produce a
nonzero field strength tensor,
\begin{equation}
 B^a_3 = F^{a12} = -\frac{\sqrt{3}}{2}\delta^{a5}gA^1_4 A^2_8,
\end{equation}
which means that the system has a uniform chromomagnetic field in it.
However, such a state would never be realized, otherwise the field
energy diverges.  To put it in another way, the vector potentials
$A_4^1$ and $A_8^2$ do not solve the Yang-Mills equations of motion,
$D_\mu F^{\mu\nu a}=0$~\cite{Larry}.  Therefore, we do not think that
the gluonic phase results from the chromomagnetic instability.

     In fact, one can reduce the field energy by making $\A_4$ and
$\A_8$ be parallel to each other.  Then the Yang-Mills action simply
vanishes.  This argument can be easily extended to more generic
$\A_\alpha$ in the three-flavor case.  We would thus reach a
conclusion that all nonvanishing $\A_\alpha$ as a result of the
chromomagnetic instability are aligned to the same direction
energetically.  Then such $\A_\alpha$ can be eliminated by a gauge
rotation of $\Delta_\eta$.  That is, the likely destination is a state
characterized by the gap parameters
$\exp[-igt^\alpha \A_\alpha\cdot\x]\cdot\Delta$ where the summation
over $\alpha$ is taken.  This is what is called the \textit{colored
crystalline phase} in Ref.~\cite{Fukushima:2005cm,Fukushima:2005fh}
and, as we have seen in (\ref{eq:act_A4}), characterized by a LOFF
ansatz beyond the single plane-wave one.  Although the difference from
the single plane-wave ansatz $\exp[-2i\q_\eta\cdot\x]\,\Delta_\eta$ is
just a color rotation, it changes the physics because $\eta$ of
$\Delta_\eta$ has the information of ``flavor'' as we have already
discussed in $\kappa^{\eta\sigma}_{\eta'\sigma'}$.
 

     In summary, based on our numerical results for the free energy as
a function of the color vector potentials, we have reached a
speculation that the chromomagnetic instability in the gapless color
superconducting region leads to the LOFF phase, meaning that the LOFF
phase is not an alternative but a destination of the instability.  In
contrast, the instability found in the 2SC (not g2SC) phase drives the
system toward a LOFF-like state which is qualitatively distinct from
the LOFF phase.


     I would like to thank Kei Iida for discussions and Krishna
Rajagopal for critical comments.  I am also grateful to Larry McLerran
for many things I learned through valuable conversations.  This work
was supported by the RIKEN BNL Research Center.


\end{document}